\documentclass[journal=nl,manuscript=article]{achemso}
\usepackage{amsmath}
\usepackage{amssymb}
\usepackage{graphicx}
\usepackage{epsfig}
\usepackage{dcolumn}
\usepackage{bm}
\usepackage{rotating}
\usepackage{multirow}
\usepackage{xspace}
\usepackage[table]{xcolor}
\usepackage[english]{babel}
\usepackage{colortbl}
\usepackage{lineno}
\usepackage{soul}

\def\bk{{\bf k}}

\def\bq{{\bf q} }

\def\bQ{{\bf Q} }

\def\<{\langle}
\def\>{\rangle}

\def\enk{ {\varepsilon}}

\let\hide\iffalse

\usepackage{braket} 
\usepackage{colortbl}
\usepackage[final]{changes}
\usepackage{amsmath}
\usepackage[utf8]{inputenc}
\usepackage{lineno}
\author{Yiming Pan}
\affiliation{Institut f\"ur Theoretische Physik und Astrophysik, Christian-Albrechts-Universit\"at zu Kiel, 24118 Kiel, Germany}
\author{Patrick-Nigel Hildebrandt}
\affiliation{Fritz Haber Institute of the Max Planck Society, Faradayweg 4-6, 14195 Berlin, Germany}

\author{Daniela Zahn}
\affiliation{Fritz Haber Institute of the Max Planck Society, Faradayweg 4-6, 14195 Berlin, Germany}
\author{Marios Zacharias}
\affiliation{Univ Rennes, INSA Rennes, CNRS, Institut FOTON - UMR 6082, F-35000 Rennes, France}
\author{Yoav William Windsor}
\affiliation{Fritz Haber Institute of the Max Planck Society, Faradayweg 4-6, 14195 Berlin, Germany}
\alsoaffiliation{Institut f\"ur Optik und Atomare Physik, Technische Universit\"at Berlin, Strasse des 17. Juni 135, 10623, Berlin, Germany}
\author{Ralph Ernstorfer}
\affiliation{Fritz Haber Institute of the Max Planck Society, Faradayweg 4-6, 14195 Berlin, Germany}
\alsoaffiliation{Institut f\"ur Optik und Atomare Physik, Technische Universit\"at Berlin, Strasse des 17. Juni 135, 10623, Berlin, Germany}
\author{Fabio Caruso}
\affiliation{Institut f\"ur Theoretische Physik und Astrophysik, Christian-Albrechts-Universit\"at zu Kiel, 24118 Kiel, Germany}
\email{caruso@physik.uni-kiel.de}
\author{H\'el\`ene Seiler}
\email{seiler@fhi-berlin.mpg.de}
\affiliation{Fritz Haber Institute of the Max Planck Society, Faradayweg 4-6, 14195 Berlin, Germany}
\alsoaffiliation{Freie Universit\"at Berlin, Arnimallee 14, 14195 Berlin, Germany}
\email{helene.seiler@fu-berlin.de}

\title{Momentum-Resolved Signatures of Carrier Screening Effects on Electron-Phonon Coupling in MoS$_2$}
\abbreviations{\replaced{UEDS}{FEDS}}

\begin{document}

\begin{abstract}

Electron-phonon coupling is central to many condensed matter phenomena. \replaced{Harnessing these effects for novel material}{Exploiting such phenomena for new} functionality in materials always involves non-equilibrium electronic states\replaced{, which in turn alter}{. In turn, these come with changes in} quasi-free-carrier density and screening. \replaced{Thus}{Therefore}, \added{gaining a fundamental }understanding \added{of} the interplay \replaced{of}{between} carrier screening and electron-phonon coupling is \replaced{essential for advancing ultrafast science.}{of fundamental importance} Prior works have mainly focused on the impact of carrier screening on electronic structure properties\replaced{. Here we investigate the non-equilibrium lattice dynamics of MoS$_2$ after a photoinduced Mott transition. The experimental data are closely reproduced by ab-initio ultrafast dynamics simulations. 
We find that the non-thermal diffuse scattering signals in the vicinity of the Bragg peaks, originating from long-wavelength phonon emission, can only be reproduced upon explicitly accounting for the screening of electron-phonon interaction introduced by the Mott transition. These results indicate the screening influences electron-phonon coupling, leading to a suppression of intravalley phonon-assisted carrier relaxation. Overall, the combined experimental and computational approach introduced here offers new prospects for exploring the influence of screening of the electron-phonon interactions and relaxation pathways in driven solids.}
{Here we combine femtosecond electron diffuse scattering and ab initio theory to show that the interactions of electrons and phonons is also altered by carrier screening in MoS$_2$. We find that screening changes the course of the non-thermal lattice dynamics by significantly slowing down intravalley scattering pathways, due to a large renormalization of the electron-phonon matrix elements for the long-wavelength phonons. Consequently, the lattice undergoes two distinct non-thermal stages before reaching quasi-thermal equilibrium 50 ps after photoexcitation. The observed modifications of the non-thermal lattice relaxation pathways demonstrate that screening can be used to control electron-phonon coupling and the ensuing lattice dynamics on ultrashort timescales. This offers prospects for advanced control schemes of material's properties, such as controlling thermal conductivity, structural phase transitions and even superconductivity.}

\end{abstract}

\begin{center}
\end{center}

\maketitle

% \section{Introduction}
Carrier screening has emerged as a powerful control knob of materials' properties, both statically and on ultrafast timescales. \cite{Elias2011} At high excitation densities, the electronic excitations generated by a femtosecond laser pulse significantly modify the dielectric environment \textit{via} carrier screening. A well-known consequence of such ultrafast photodoping is bandgap renormalization, an effect which has been investigated for decades as observed in many materials, ranging from semiconductors to quantum materials.\cite{Shah1977, Reynolds2000, Chernikov2015, Chernikov2015_2, Smallwood2012,MorHerzog2017, DendzikXian2020,CalatiLi2023} 
Recently, it was demonstrated that dynamical screening can also be the origin of emergent electronic phases that are only accessible out-of-the equilibrium.\cite{HuberLin2024, Beaulieu2021,DuanXiaHuang23} For example, carrier screening was predicted to modify the effective electronic correlations dynamically,\cite{dalatorre2021,Tancogne-Dejean2018} and leads to an ultrafast Lifschitz transition in MoTe$_2$ as observed in photoemission experiments.\cite{Beaulieu2021} 
In semiconducting transition metal dichalcogenides (TMDCs), carrier screening leads to the decrease of the attractive interactions of the strongly bound excitons and a reduction of the exciton binding energy.\cite{GaoYang2016,LiangYang2015} 
Beyond certain carrier densities, a Mott transition where the excitons are quenched and dissociate into free electron and hole plasma occurs.\cite{YuBataller2019, Pogna2016,Chernikov2015_2, Steinhoff2017, Bataller2019, Shah1977,Perfetto2021,Liu2019}  Overall, these examples demonstrate that femtosecond laser pulses offer a way to alter a material’s response \textit{via} dynamical screening on ultrafast timescales.

Prior works have mostly focused on the impact of carrier screening on the properties of electronic structures. Much less is known regarding the effects of carrier screening on the lattice interactions. Recently, dielectric screening from the substrate was shown to significantly alter the phonon dynamics in monolayer MoS$_2$. \cite{Britt2022}
However, to the best of our knowledge, experimental evidence of how carrier screening within a material modifies electron-phonon coupling has remained scarce. From the theory side, the renormalization of electron-phonon coupling (EPC) matrix elements due to screening effects suggests that the lattice dynamics can be affected by changes in the dielectric environment.\cite{Verdi2017,Riley2018,Britt2022,Maldonado2020} Furthermore, 
the screening introduced by the electron-hole plasma is particularly effective for lattice vibrations which create macroscopic electric fields of wavelength longer than the screening length,\cite{Macheda2022,Macheda2023, SohierGibertini2021} modifying the interactions between electron and lattice and possibly allowing access to hidden quantum phases of the lattice. \cite{Peng2020, Paillard2019} Understanding how electron-phonon coupling and the lattice dynamics are impacted by carrier screening is therefore relevant both at the fundamental level as well as in view of devising new lattice control schemes.  

Here we induce a Mott transition in bulk MoS$_2$ with an ultrashort laser pulse and directly investigate the lattice dynamics in momentum space using a combination of femtosecond electron diffuse scattering (FEDS) and first-principle calculations which include free-carrier screening to the EPC matrix elements. Working with bulk MoS$_2$ enables us to avoid screening effects from the substrate \cite{Britt2022} and to focus on carrier screening that is intrinsic to the material.
\replaced{Surprisingly, our data reveal that lattice relaxation involves two qualitatively distinct non-thermal phonon populations. Ultrafast dynamics simulations using unscreened EPC matrix elements successfully capture the short-wavelength phonon emissions and the lattice relaxation for time delays beyond 5 ps. On shorter timescales, however, the diffuse scattering signals in the vicinity of the Bragg peaks can only be reproduced upon explicitly accounting for the screening of the electron-phonon interactions in ab-initio calculations. These finding indicate that screening affects primarily long-wavelength phonons around the $\Gamma$ point and leads to a suppression of intravalley scattering pathways. Overall, our results suggest that carrier screening profoundly alters materials' electron-phonon coupling and lattice relaxation pathways, providing access to hidden dynamical regimes.}{We find that carrier screening modifies the lattice dynamics on the several picosecond timescale due to a drastic suppression in the emission of long-wavelength phonons. As a result, the photogenerated electrons and holes couple preferentially to short-wavelength phonons at the Brillouin zone edge, generating a highly non-thermal distribution of phonons over 500 fs. 
Surprisingly, before reaching a quasi-thermal state over timescales of 50~ps, the lattice dynamics transition towards a second non-thermal state that persists for several picoseconds, disclosing multi-step thermalization dynamics.
Overall, our results reveal that carrier screening profoundly alters materials' electron-phonon coupling and lattice relaxation pathways, providing access to hidden dynamical regimes.} Given that electron-phonon coupling and screening are ubiquitous in condensed matter, our findings are generally applicable. Exploring such EPC renormalization in the future offers new opportunities for devising schemes to control lattice properties on ultrafast timescales, which ultimately could lead to new functionalities in quantum technologies and opto-electronic devices.
 
% \section{Results and Discussion}
\section{Results}
\begin{figure*}[ht!]
\centering
\includegraphics[width=1\textwidth]{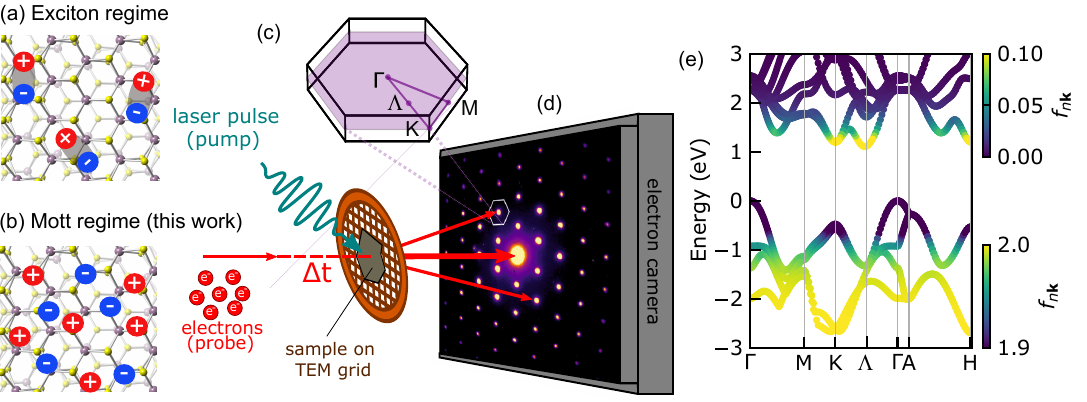}
\caption{Crystal structure of multilayer MoS$_2$ and schematic illustration of (a) exciton regime at low carrier excitation densities (b) Mott regime at high carrier excitation densities (\textgreater 1.5 $\cdot 10^{13}$ cm$^{-2}$). (c) Brillouin zone (BZ) and high symmetry points of bulk MoS$_2$. (d) Schematic illustration of the FED experiment. (e) Calculated electron band structure of bulk MoS$_2$ and initial distribution of photo-excited carriers superposed as a color coding .}
\label{fig:1}
\end{figure*}

The layered hexagonal structure of bulk MoS$_2$ is shown from the top in Fig.~\ref{fig:1}(a-b), whereas the 3D Brillouin zone (BZ) is displayed in Fig.~\ref{fig:1}(c) along with high-symmetry points. In our diffraction experiment, we obtain a direct view of the reciprocal lattice for momenta within the $\Gamma$-K-M plane in the BZ (shaded purple plane in Fig.~\ref{fig:1}(c)). \added{Each Bragg peak acts like the center of a first BZ, located at the $\Gamma$ point.}
An exemplary diffraction pattern obtained on a mechanically exfoliated MoS$_2$ flake of thickness $\simeq$ 35 nm is shown in Fig.~\ref{fig:1}(d), which illustrates the basic idea of an FED experiment.\cite{2015Wald} Here we employ $\approx$50~fs, 2.1 eV pump pulses to excite the MoS$_2$ flake.  
We estimate our excitation density to be in the range of $1.29-1.69 \cdot 10^{14}$ cm$^{-2}$, well into the Mott regime.\cite{Chernikov2015,Chernikov2015_2, Steinhoff2017, Bataller2019, Shah1977} (see section 1 of the Supplementary Information (SI)\cite{supp1} ). 
Therefore, optical excitation quickly leads to the formation of an electron-hole plasma and exciton effects are suppressed by screening, as depicted in Fig.~\ref{fig:1}(a-b). \replaced{Following the generation of electrons and holes, electron-phonon coupling transfers energy from the photo-excited carriers to the crystal lattice. Because electron-phonon coupling varies for different modes, energy can be preferentially deposited in strongly-coupled modes, leading to selected phonons with larger occupation number. In such case, the lattice cannot be described by Bose-Einstein statistics, resulting in a non-thermal phonon distribution. FEDS is ideally suited to probe non-thermal phonons due to its momentum information. Following the optical pump, the non-equilibrium lattice dynamics in the sample is probed by a 150~fs bunch of electrons which diffracts off the lattice. The FEDS signals vary linearly with phonon populations, and their transient changes directly reflect the time evolution of the non-thermal phonon populations at different wavevectors in the BZ. For instance, long-wavelength phonons affect the FEDS signals in the vicinity of the Bragg peaks close to $\Gamma$, while short wavelength phonons induce changes at $\rm K$ or $\rm M$. More details about FEDS can be found in Ref. \citenum{2015Wald, Stern2018, Waldecker2017, BP_Seiler, Zacharias2021-1}.}{Following the optical pump, we probe the non-equilibrium lattice dynamics in the sample with a femtosecond bunch of electrons with an estimated duration of 150~fs.} All measurements are carried out at room temperature. 

Before discussing the lattice dynamics, we consider the initial populations of electrons and holes in more detail. Fig.~\ref{fig:1}(e) displays the electronic band structure of bulk MoS$_2$ calculated with density functional theory (DFT).\cite{Giannozzi2017} In contrast to monolayer MoS$_2$, bulk MoS$_2$ is a semiconductor with an indirect bandgap of 1.29 eV between $\Gamma$ and its conduction band minimum (CBM) located at $\Lambda$, around the midpoint between $\Gamma$ and K. \cite{Padilha2014,Cheiwchanchamnangij2012} 
Within several tens of femtoseconds, the excited carriers thermalize within conduction bands and valence bands \textit{via} electron-electron scatterings.\cite{Tanaka2003, Nie2014,Puppin2022}The superimposed color code on the band structure shows the initial ($t = 0$) thermalized distribution of photoexcited carriers $f{_{n\bk}}$ in the conduction and valence bands. 
These are simulated with Fermi-Dirac distributions with a hot carrier temperature (4000 K) and different chemical potentials for electrons and holes to fix the carrier density to the experimental conditions. \replaced{The electronic temperature is chosen such that the final lattice temperature in the simulations match the estimated temperature rise in the experiments of $\Delta T = 70 ~\rm K$.}{We describe this method \added{in detail} in section 1 of SI.}  The initial distribution of electrons and holes shows that in conduction bands primarily the K and $\Lambda$ pockets are occupied by electrons and in valence bands the K and $\Gamma$ are populated by holes after photo-excitation\cite{Bertoni2016}. Initial distribution with other carrier temperatures lead to \added{qualitatively} similar results, \added{with different final lattice temperatures.} \added{We discuss these results} \replaced{in sections 2 and 3}{reported} of \added{the} SI.

\begin{figure*}[ht!]
\centering
\includegraphics[width=1\textwidth]{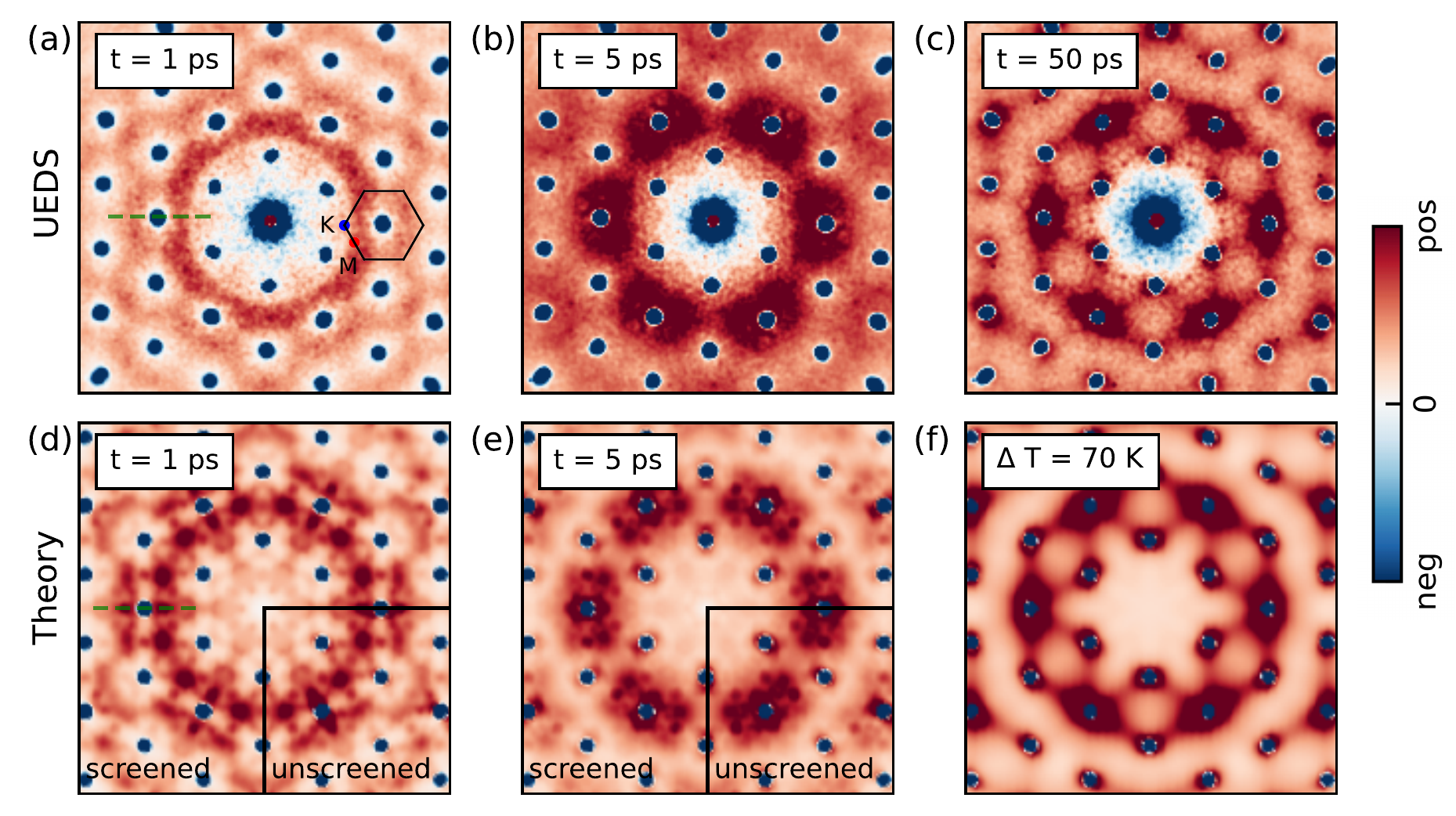}
\caption{Differential diffuse scattering signals $\Delta I(\bQ, t) = I(\bQ, t)-I(\bQ, t\leq t_0)$ at pump-probe delays of (a) 1 ps (b) 5 ps and (c) 50 ps from FEDS experiments. Simulated diffuse scattering maps at pump-probe delays of (d) 1 ps with screened EPC matrix element and unscreened EPC in the 4th quadrant of the map (e) 5 ps with screened EPC matrix element \added{and unscreened EPC in the 4th quadrant of the map}. (f) Thermal diffuse scattering pattern, $\Delta I(\bQ) =  I(\bQ, T= 370 $\rm \  K$) -  I(\bQ, T= 300 $\rm \ K$)$, where 370 K is the estimated final temperature and 300 K is the initial lattice temperature. 
%\replaced{ One-dimensional cuts at 1 ps (along green dashed lines in panels (a) and (d)) along the (110) Bragg peak for unscreened (red), screened (blue) and experiments (dashed black). (h) Same but for 2 ps. }
\deleted{(g) Zoom-in close to the Brillouin zone around Bragg peak (110) for the simulated diffuse scattering map with unscreened EPC matrix element (g), screened EPC matrix element (h) and experimental data (i) at 1 ps. }}
\label{fig:2}
\end{figure*}

A detailed view of phonon dynamics across the BZ is obtained by inspecting inelastic scattering signals between the Bragg peaks \textit{via} FEDS.\cite{Waldecker2017, Stern2018, Cotret2019, BP_Seiler, Britt2022, BrittCaruso2024, Maldonado2020}. 
Details are discussed in section \replaced{4}{3} of the SI. The Bragg peak dynamics are discussed in section \replaced{5}{4} of the SI. An overview of the FEDS signals from the experiments is shown in Fig.~\ref{fig:2}. Here, the differential diffuse scattering signals $\Delta I(\bQ, t) $ are obtained by substracting the diffraction intensity at pump-probe time delay $t$ from the equilibrium diffraction pattern $I(\bQ, t\leq t_0)$ before excitation. In panel (a)-(c) we present $\Delta I(\bQ, t) $ for pump-probe time delays of 1 ps, 5 ps and 50 ps.These data demonstrate qualitative changes in the distribution of the inelastic signals as pump-probe delay increases, reflecting different phonon populations at different times. Strikingly, these data display signatures of two distinct non-thermal phonon distributions. The inelastic scattering distribution at 1 ps qualitatively differs from the one at 5 ps, which in turn differs from the distribution at 50 ps. 

To gain further microscopic insights into the coupled electron-lattice dynamics seen in the experiments, we perform ab-initio calculations of the non-equilibrium diffuse scattering. We simulate the real-time dynamics of the coupled electron and phonon systems by solving the time-dependent Boltzmann Equation (TDBE).\cite{Allen1987,Sadasivam2017,Caruso2021,Caruso2022,Tong2021,Maldonado2017,PanCaruso2023,PanCaruso2024} We use the initial carrier distribution as discussed in Fig.~\ref{fig:1}(e) and initial phonon population with a thermal distribution at 300 K. These results are then used as inputs for computations of the all-phonon structure factor in non-equilibrium. \cite{Zacharias2021-1, Zacharias2021-2, PENG1999625} This approach enables direct comparison of the theoretical and experimental data.\cite{BP_Seiler,Britt2022,Maldonado2020} The simulation results are shown in Fig.~\ref{fig:2}(d-f), and a movie with the comparison between simulations and experiments at all pump-probe delays is available as supporting information. 

\replaced{We perform simulations with and without carrier screening effects arising}{In our simulations, we take into account carrier screening effects arising} from the Mott transition induced by the intense driving field in the FEDS experiments. The influence of photoexcited carriers on phonon dynamics is included by screening the EPC matrix elements\cite{Macheda2022,Macheda2023,Smejkal2021,Verdi2017,Riley2018,Kandolf2022,Caruso2016} \textit{via} $\tilde{g}_{mn\nu} (\bk, \bq) = \epsilon^{-1}_{\rm MT}(\bq) g_{mn\nu} (\bk, \bq)$, where the $g_{mn\nu} (\bk, \bq)$ are the EPC matrix elements without carrier doping obtained from density-functional perturbation theory.\cite{Baroni_2001}
$\epsilon^{-1}_{\rm MT}(\bq)$ accounts for additional screening components to phonons of wave vector $\bq$,  arising from the Mott transition and the photoexcited free carriers. In the independent-particle approximation:

\begin{align}\label{IPA}
\epsilon_{\rm MT}(\bq) = 1 - \frac{2}{\Omega N_{\bf k}}\frac{4\pi e^2}{|{\bf q}|^2 \epsilon_{\rm undop}({\bf q})} \sum_{mm'{\bf k}} \frac{\delta f_{m {\bf k}} - \delta f_{m' {\bf k+q}}}{\varepsilon_{m {\bf k}}- \varepsilon_{m' {\bf k+q}}} |\langle{u_{m{\bf k}}}|{u_{m'{\bf k+q}}}\rangle|^2 
\end{align}
where $\delta f_{m {\bf k}}$ accounts for the change of carrier occupations due to optical excitation. The electron energies $\enk_{m\bk}$ and the periodic part of the Bloch eigenvectors $\ket{u_{m \bk}}$ are obtained from DFT and Wannier interpolation techniques.\cite{pizzi2020wannier90,Marzari2012} $\epsilon_{\rm undop}$ indicates the dielectric function of the undoped MoS$_2$. 
Equation~\eqref{IPA} and its numerical evaluation are discussed in section \replaced{6}{5} of the SI.  

\begin{figure*}[ht!]
\centering
\includegraphics[width=1\textwidth]{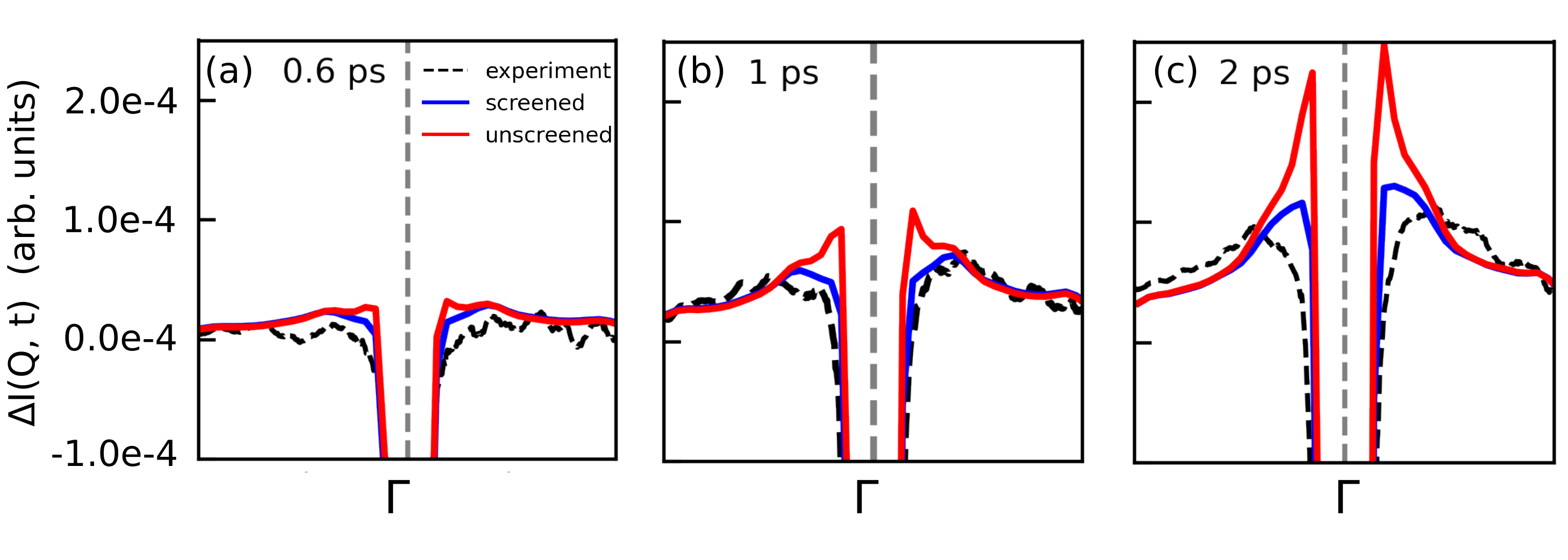}
\caption{\added{Transient diffuse scattering intensity at (a) 0.6 ps, (b) 1.0 ps and (c) 2.0 ps in the vicinity of the (110) Bragg peak for unscreened (red) and screened (blue) simulations, and experiments (dashed black). The phonon momentum on the x axis follows the path marked by the dashed green lines in Fig.~\ref{fig:2} (a) and (d).}}
\label{fig:3}
\end{figure*}

The comparison of experimental and simulated data in Fig.~\ref{fig:2} shows that the ab initio simulations successfully capture the two-stage lattice relaxation observed in the experiments. Finer structure features are seen in the theory maps, which we cannot capture in our experiments due to signal-to-noise limitations. The influence of carrier screening on the diffuse scattering signals can be determined by comparing the diffuse scattering maps computed with the non-equilibrium phonon populations obtained with screened and unscreened EPC matrix elements. In Fig.~\ref{fig:2}(d), the forth quadrant is the result with unscreened EPC and the other three quadrants are the results with screened EPC. \replaced{Overall, the differences between the screened and unscreened cases are mainly visible around the Bragg peak area, while higher momenta of the BZ remain largely unaffected. Furthemore, we observe that the differences between unscreened and screened cases are largely reduced by 5 ps, see Fig.~\ref{fig:2}(e). To qualitatively assess the influence of screening on the emission of long-wavelength phonons, we report in Fig.~\ref{fig:3} the FEDS intensity along a path in reciprocal space at different time delays, marked by the green dashed line in Fig.~\ref{fig:2}(a) and (d). In the absence of carrier screening, simulations predict a sharp rise of the FEDS intensity close to the Bragg peak, due to the large phonon populations emitted through intravalley relaxation pathways. Upon accounting for screening of the electron-phonon coupling, conversely, the emission of long-wavelength phonons is strongly suppressed, leading to a substantiated reduction in the FEDS intensity in the vicinity of the Bragg peaks. Overall, the inclusion of screening is crucial to qualitatively reproducing the experimental trend.}{Fig.~\ref{fig:2}(g)-(i) display a zoom-in of the BZ around Bragg peak (110) for experiments and calculations
. A comparison of the screened and unscreened simulated signals show a suppression of the low wavector phonons around $\Gamma$ in the screened case. This matches well with the experimental observation of the white ring around the Bragg peaks on the 1 ps map, and in Fig.  \ref{fig:3}(i).} These results \replaced{open new prospects for exploring the influence of}{demonstrate that} screening \deleted{effects} on the lattice dynamics \replaced{through the combination of FEDS experiments and ab-initio calculations}{can be identified \textit{via} qualitative differences in the diffuse scattering fingerprints} \added{in the vicinity of the Bragg peaks}.

\begin{figure*}[ht!]
\centering
\includegraphics[width=1\textwidth]{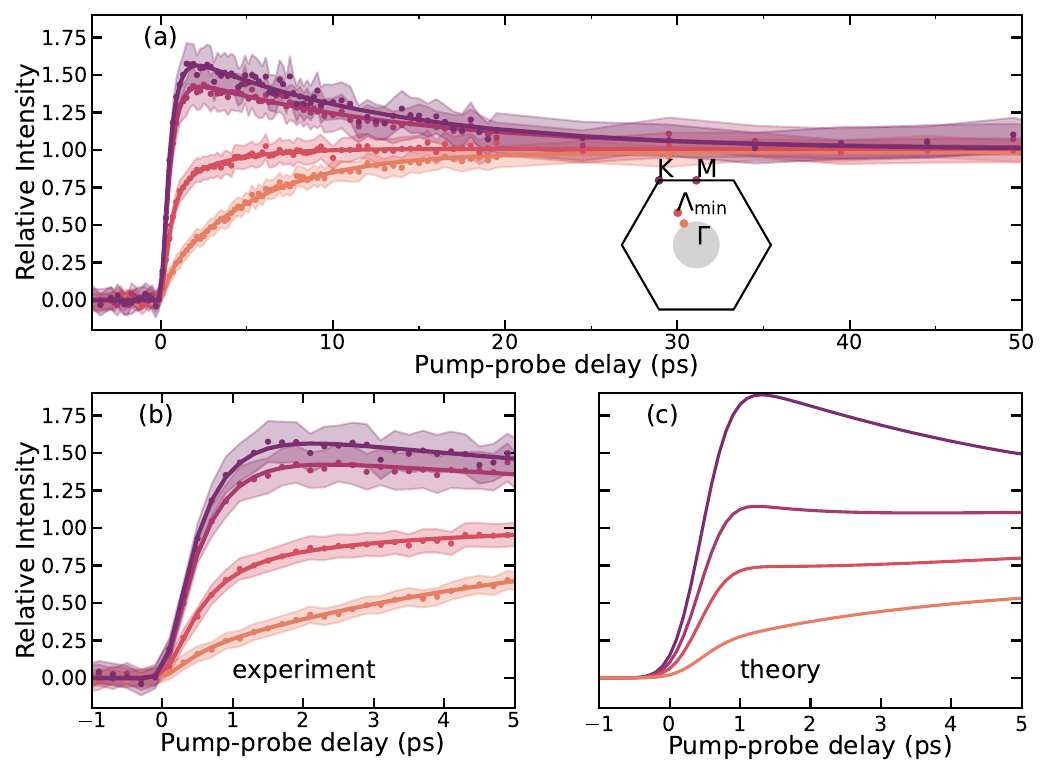}
\caption{ (a) Time resolved relative differential diffuse scattering intensity at selected $\bq$ points along the $\Gamma$-$\Lambda$-K high-symmetry path \added{as well as for the M point}, indicated as colored dots on the BZ in the inset. The shaded areas correspond to the standard error of the mean signal over multiple delay scans. The relative intensities are normalized to the late time signals at 50 ps and are fitted with a bi-exponential function (continuous lines). (b) Same as panel (a) for the first 5 ps. (c) Simulated relative differential diffuse scattering intensities for the same $\bq$ points as in (a) and (b), matching colors. The simulated traces were convolved with a Gaussian corresponding to the finite IRF of the experiments. }
\label{fig:4}
\end{figure*}

Quantitative information on the non-equilibrium phonon dynamics can be retrieved \textit{via} a momentum-resolved analysis of the diffuse scattering signals. 
It is desirable to extract the relative intensity
at specific momenta. Fig.~\ref{fig:4} displays relative diffuse scattering intensity
at selected q-points in the BZ (see section \replaced{7}{6} of the SI for more detail about the extraction of the signals). The fastest and strongest amplitude rise are observed for M and K phonons, with time constants of $0.43 \pm 0.02$ ps and $0.55 \pm 0.02$, respectively, matching well with the fast time constant measured in the Bragg peak dynamics. \added{From our simulations, we retrieve consistent relative intensity amplitudes. We find that the M point dynamics is 
%$x \pm pm$
0.12 ps, while the K point dynamics is %$y \pm pm$
0.18 ps. In the experiments, the instrument response function (IRF) can only be estimated and not directly measured, which may cause some variations in the extracted time constants and may explain the faster time constants observed in the simulations.} The rise is followed by a similar decay with a time constant of $11.8 \pm 0.8$ ps and $11.6 \pm 0.7$ ps for M and K phonons, respectively. Following the fast rise of K and M phonons, the phonon dynamics at $\Lambda$ can be fitted with two rising exponentials, with time constants of $0.6 \pm 0.1$ and $3.1 \pm 0.5$ ps. We observe that the closer we approach $\Gamma$, the slower the phonon population rises. The experimental traces shown in Fig.~\ref{fig:4}(b) can be directly compared with the temporal evolution of the relative intensities from our ab initio computations, shown in panel (c). We make quantitative analysis of the nonequilibrium lattice dynamics by tracking the relative intensity as a function of delay time for 3 points along the $\rm \Gamma-K$ high-symmetry path (see section \replaced{7}{6} of the SI) as well as at the M point \deleted{(see section 7 of the SI)}. The comparison of simulated and experimental relative intensities shown in Fig.~\ref{fig:4}(b-c) demonstrate an agreement between the FEDS experiments and TDBE simulations.

\begin{figure*}[ht!]
\centering
\includegraphics[width=1\textwidth]
{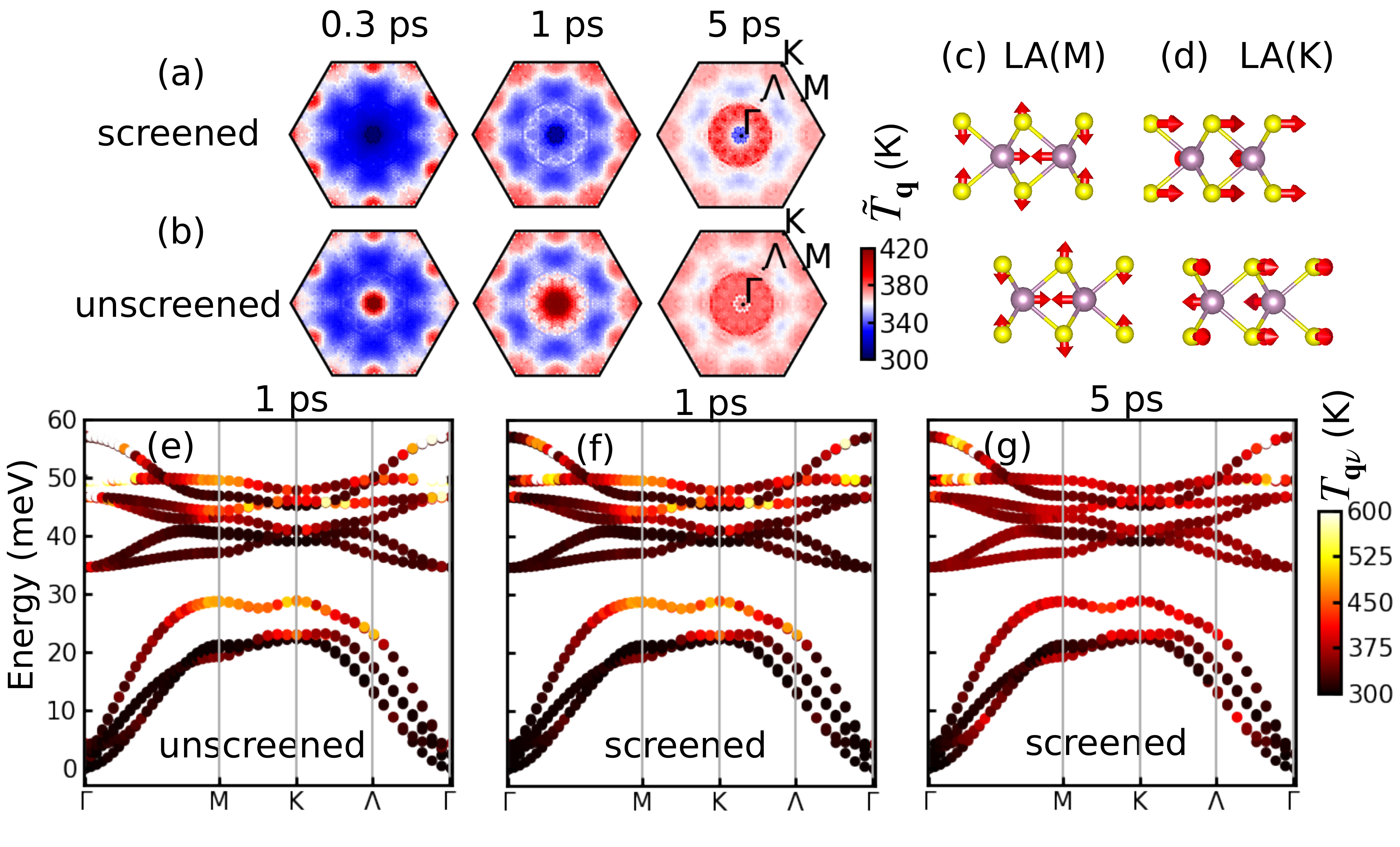}
\caption{Effective phonon temperature $\tilde{T}_{\bq} $ at 0.3 ps, 1 ps and 5 ps simulated with screened (a) and unscreened (b) EPC matrix elements. (c) The eigenmode of LA(M) and (d) LA(K) phonons. (e) The branch- and momentum-resolved phonon temperature $T_{\bq\nu}$ superposed on the phonon dispersion along the high-symmetry path $\Gamma$-M-K-$\Gamma$ at 1 ps calculated with unscreened, 1 ps (f) and 5 ps (g) with screened EPC matrix elements.}
\label{fig:5}
\end{figure*}

Next we show that carrier screening profoundly influences the non-thermal relaxation pathways, by suppressing specific phonon-assisted recombination channels. \replaced{These mechanisms modify the two-stage non-thermal phonon populations observed in our FEDS experiments.}{These mechanisms are at the origin of the two-stage non-thermal phonon populations observed in our FEDS experiments.}
\replaced{Fig.~\ref{fig:5}}{Fig.~4} summarizes the non-equilibrium phonon dynamics obtained from our simulations. As an indicator of the non-thermal phonon populations established out of equilibrium,  we introduce the  effective phonon temperature, $\tilde{T}_{\bq} = N_{ph}^{-1}\sum_{\nu} T_{\bq \nu}$, where $T_{\bq \nu} = \hbar\omega_{\bq\nu} \{k_B {\rm ln} [1+n_{\bq \nu}^{-1}] \}^{-1}$, and $n_{\bq \nu}$ and $\omega_{\bq \nu}$ are the phonon distribution and frequencies, respectively. 
\replaced{Fig.~\ref{fig:5}}{Fig.~4}(a) and (b) report values of $\tilde{T}_{\bq}$ for momenta in the BZ at time delays of $t =$ 0.3, 1 and 5 ps. 
The results in panel (a) have been obtained by including carrier screening of the EPC induced by the photoinduced Mott transition \textit{via} Eq.~\eqref{IPA}, whereas these effects are omitted in panel (b). 
These two regimes reveal large qualitative differences in the non-equilibrium dynamics of long-wavelength phonons\added{, in particular $<$ 5 ps}. 
In the unscreened case, the population of phonons close to $\Gamma$ increases rapidly within 1~ps, revealing carrier relaxation pathways mediated by intravalley electron scattering. 
Conversely, in the presence of screening, intravalley scattering is inhibited, resulting in the suppression of phonon emission around $\Gamma$. 
Long-wavelength phonons are only subsequently populated \textit{via} phonon-phonon scattering on a timescale of 5~ps. The differences between screened and unscreened cases are further manifested by the mode- and momentum-resolved effective phonon temperatures $T_{\bq \nu}$, superimposed to the phonon dispersion curve in \replaced{Fig.~\ref{fig:5}}{Fig.~4} (e) and (f) for $t$ = 1 ps. 
These results reveal that screening affects primarily the dynamics of long-wavelength phonons, which is critical to retrieve the diffuse scattering fingerprints close to the Bragg peaks, as demonstrated by comparing \replaced{the 1D cuts shown in Fig.~\ref{fig:3}(a)-(c) for the unscreened, screened and experimental cases.}{the differential diffuse scattering signals in Fig.~2(g)-(i).} The screening affects less the emission of LA phonons at M and K points, which contributes to the diffuse scattering due to the in-plane atomic motions for these two modes, as demonstrated by their eigenvectors in \replaced{Fig.~\ref{fig:5}}{Fig.~\ref{fig:2}}(c) and (d).

\begin{figure*}[ht!]
\centering
\includegraphics[width=1\textwidth]{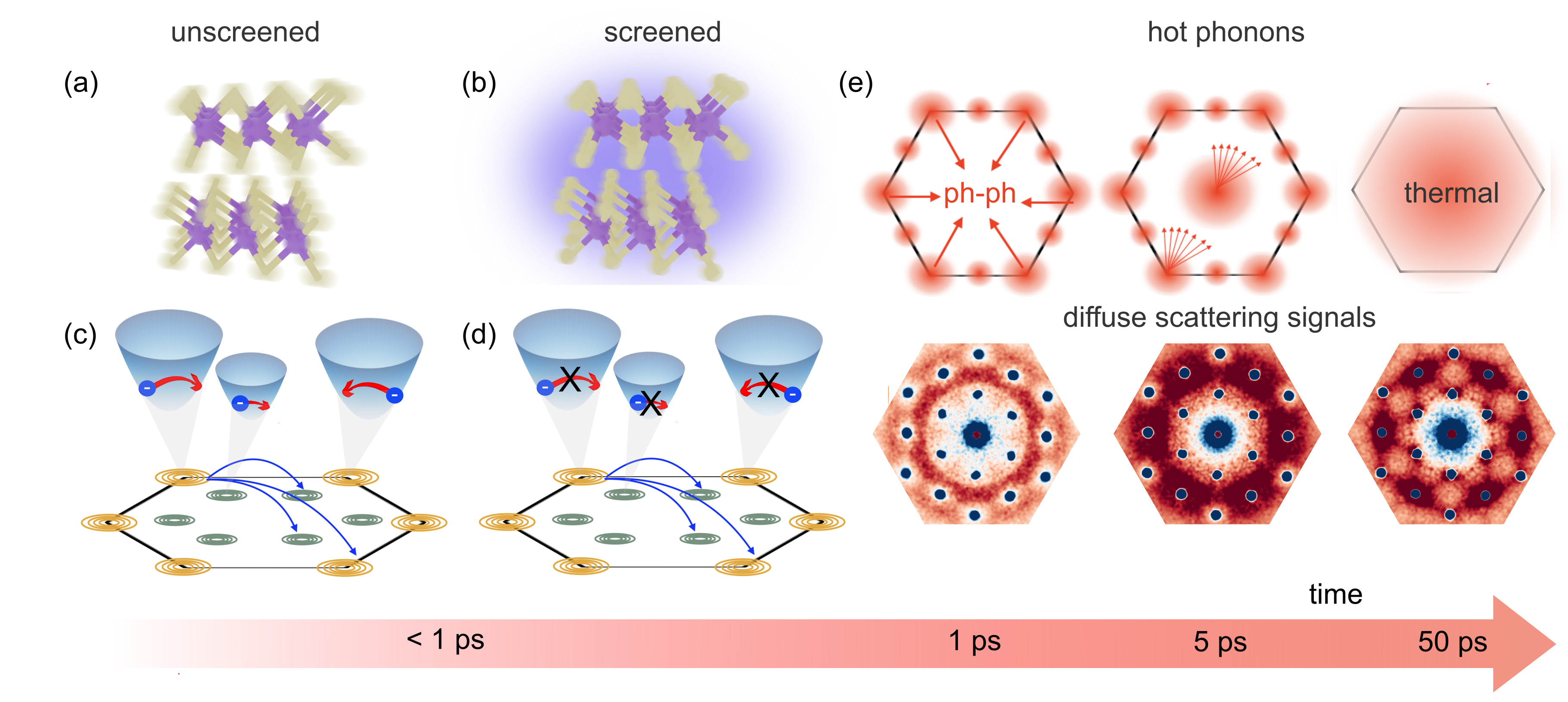}
\caption{Impact of carrier screening on electron-phonon coupling and resulting modifications of the non-thermal lattice relaxation pathways.(a)-(b) Phonon emission close to the BZ center is strongly reduced as a result of carrier screening. This is illustrated for an examplary high-frequency, long wavelength optical phonon mode, which shows decreased fluctuations of atomic motions in the screened case. (c)-(d) screening suppresses intravalley scattering of carriers, while intervalley scattering pathways are preserved. (e) Phonons are first excited at the BZ edge on the 1 ps timescale. These phonons are then scattered to the center of BZ on the 5 ps timescale. Finally, a quasi-thermalized state of the lattice is reached at 50 ps. The diffuse scattering signals corresponding to the hot phonons populations are shown below.}
\label{fig:6}
\end{figure*}

Bringing the results of Figs.~\ref{fig:2}-\ref{fig:6} together, we obtain a microscopic picture of how the momentum-dependent renormalization of the EPC due to screening changes the course of non-thermal phonon relaxation pathways in MoS$_2$. We summarize the thermalization steps in \replaced{Fig.~\ref{fig:6}}{Fig.~\ref{fig:5}}. We photo-induce a Mott transition in MoS$_2$ using an ultrashort laser pulse. Electron-phonon intravalley scattering with low-momenta phonons (in the $\Gamma$ - 1/3K range) are strongly suppressed due to carrier screening, see the sketches in \replaced{Fig~\ref{fig:6}}{Fig.~\ref{fig:5}}(a-b). As a result, the first stage of the electron-phonon dynamics is dominated by intervalley scattering processes, \added{which are unmodified by screening,} see \replaced{Fig.~\ref{fig:6}}{Fig.~\ref{fig:5}}(c-d). Inspecting the multivalley electronic band structure in Fig. \ref{fig:1}(e), one can expect K-K intervalley scattering in the conduction bands and $\Gamma$-K scattering in the valence bands to give rise to phonon emission at the K points of the BZ. Furthermore, K-$\Lambda$ scattering processes in the conduction band are expected to produce significant phonon emission at the M points of the BZ, whereas the scattering between $\Lambda$ points should give rise to phonon emission at K, $\Lambda$ and M. This is fully consistent with the experimental observations seen in Fig.~\ref{fig:2}.

In the second stage of the phonon dynamics, phonon-phonon interactions lead to a more isotropic momentum-resolved temperature at 5 ps, as demonstrated in Fig.~\ref{fig:6}(a) and (g). Nevertheless, the lattice remains in a non-thermal state\added{, with minor differences persisting between the screened and unscreened cases}. \deleted{The differences between screened and unscreened cases in Fig.~5(a-b) shows that even several picoseconds after photo-excitation, consequences of the EPC renormalization due to screening remain visible in the data.}In this second stage, we attribute the origin of the non-thermal state of the lattice to two processes: (i) slower electron-phonon scattering of low momenta phonons due to strong screening of the EPC for such phonons and (ii) slower phonon-phonon scattering rates of the acoustic branches in comparison with that of optical phonons, as illustrated by the anharmonic phonon scattering rate in section 8 of SI. A visible increase of phonon population around the $\Gamma$ point is seen, but the TA and LA branches remain strongly out-of-equilibrium in momentum space. The thermalization limited by acoustic phonons seems to be a common feature in layered 2D materials, as investigated in recent works.\cite{BP_Seiler,Kurtz2024,PanCaruso2023, PanCaruso2024} Finally, at around 50 ps, the lattice reaches thermalization at an elevated temperature, seen in Figs.~\ref{fig:2}(c) and (f). These steps are illustrated schematically in \replaced{Fig.~\ref{fig:6}}{Fig.~\ref{fig:5}}(e).
\section{Conclusions}
Our study \added{introduces a momentum-resolved approach to investigate the effect of carrier screening on electron-phonon coupling, and }provides \added{first} experimental evidence that large modifications of electron-phonon coupling are induced \textit{via} photoinduced screening. 
% \added{We hope our approach will be further improved in future studies as FEDS matures and signal-to-noise in experiments improve, enabling fluence-dependent measurements. Better beam coherence will be essential to reach closer to the $\Gamma$ point, where the differences between screened and unscreened should become most pronounced.} 
\added{Future advancements in FEDS will open new avenues for exploring screening dynamics. Enhanced beam coherence and improved signal-to-noise ratio in experiments will be particularly valuable in accessing regions closer to the $\Gamma$ point, where the contrast between screened and unscreened scenarios is expected to be most pronounced. These developments will further advance our understanding of screening, electron-phonon interactions and their interplay in ultrafast phenomena.}
Electron-phonon coupling and carrier screening are both ubiquitous in condensed matter systems, thus our findings can be generalized beyond MoS$_2$. We expect carrier screening to similarly renormalize electron-phonon coupling for the long-wavelength phonons in other photo-excited materials at high excitation densities. Consequently, ultrafast photodoping can be used as a versatile tool to alter the  non-thermal lattice relaxation pathways in materials, and our approach can be applied to diverse material platforms, ranging from semiconductors to quantum materials. We envision that control of electron-phonon interactions \textit{via} carrier screening will give rise to advanced control schemes of materials' properties, to reach hidden quantum phases that are not accessible in equilibrium or tuning properties such as thermal or electrical conductivity on the ultrafast timescale.

\section{Methods}

\subsection{Computational details}

The ground-state properties of bulk MoS$_2$ are obtained from Kohn-Sham density-functional theory (DFT) as implemented in the plane-wave pseudopotential code {\tt Quantum Espresso}.\cite{Giannozzi2017} We used norm-conserving Hartwigsen-Goedecker-Hutter pseudopotential\cite{Hartwigsen1998} with Perdew-Burke-Ernzerhof generalized gradient approximation (GGA-PBE) to the exchange-correlation functional.\cite{GGA_Pedrew_1996} All calculations employed the DFT-relaxed crystal structure. The BZ is sampled with 12 $\times$ 12 $\times$ 4 mesh. The phonons are calculated with a $\bq$-grid of 5 $\times$ 5 $\times$ 2 in the BZ based on density-functional perturbation theory (DFPT).\cite{Baroni_2001} Spin-orbital coupling is neglected in our calculations.

The electron-phonon coupling matrix elements are calculated within the {\tt EPW} code,\cite{bib:epw,Giustino2007} which uses {\tt Wannier90}\cite{pizzi2020wannier90} as a module. The wave functions from DFT calculations are projected onto 22 maximally-localized Wannier functions (MLWF)\cite{Marzari2012,pizzi2020wannier90} using $d$ orbitals of Mo and $p$ orbitals of S as initial projectors. The electron energies, phonon frequencies, and electron-phonon coupling matrix elements are interpolated on a 48 $\times$ 48 $\times$ 10 Monkhorst-Pack homogeneous mesh for both $\bk$ and $\bq$ points using MLWF. We used a smearing parameter of 3 meV for the calculation of the electron-phonon collision integral and 0.05 meV for the evaluation of the phonon-phonon collision integral. The third-order force constants --  required for the evaluation of momentum- and mode-resolved phonon-phonon relaxation times -- is obtained from finite differences using the {\tt third-order.py} utility of the {\tt shengBTE} code.\cite{LI20141747} Calculations are performed  on a 3 $\times$ 3 $\times$ 2 supercell. Ab initio simulations of the ultrafast electron-phonon dynamics are based on the TDBE, which we have implemented in the {\tt EPW} code. The time derivative is computed using the second-order Runge-Kutta (Heun's) method with a time step of 2 fs. The details of the collision integrals and the implementations are provided elsewhere.\cite{Caruso2021, Caruso2022}

\begin{acknowledgement}
The authors gratefully acknowledge the
computing time provided by the high-performance computer Lichtenberg at the NHR Centers NHR4CES at TU
Darmstadt (Project p0021280). This work received funding from the Deutsche Forschungsgemeinschaft (DFG) within Transregio TRR 227 Ultrafast Spin Dynamics (Projects A10 and B11). Funding was also received from the Max Planck Society and the European Research Council (ERC) under the European Union’s Horizon 2020 research and innovation program (Grant Agreement No. ERC-2015-CoG-682843).  Y. P acknowledges funding by the DFG -- project number 443988403. H. S. acknowledges support from the Free University of Berlin and from the Swiss National Science Foundation under Grant No. P2SKP2.184100. M.Z. was funded by the European Union (project ULTRA-2DPK / HORIZON-MSCA-2022-PF-01 / Grant Agreement No. 101106654).
\end{acknowledgement}

\section{Author contributions}
H.S., F.C. and R.E. conceived the research project. Y.P. and F.C. developed the theory. Simulations and numerical implementations were conducted by Y.P. and M.Z. Y.P implemented the TDBE in EPW code and M.Z. developed the code for structure factor simulations.  H.S. prepared the sample. H.S. and D.Z. performed the experiments. P.N.H. developed data analysis tools and analysed the experimental data, under the guidance of H.S. H.S., D.Z., and Y.W.W. participated in maintaining and running the experimental apparatus. Y.P, H.S. and F.C. wrote the manuscript. Other co-authors commented on the manuscript. R.E. provided funding for the experiments.

\bibliography{bibliography}
\end{document}